\documentclass[final,english]{bullsrsl}[2022/06/15]

\usepackage[latin1]{inputenc}
\usepackage[T1]{fontenc}

\usepackage{natbib} 
\usepackage{graphicx}

\newcommand{\kms}{km s$^{-1}$}

\newcommand{{\lrsp}}{{\it LRSP}}

\begin{document}
\title{Observational Characteristics of solar EUV waves}

\author[affil={1},corresponding]{Ramesh}{Chandra}
\author[affil={1}]{Pooja}{Devi}
\author[affil={2}]{P. F.}{Chen}
\author[affil={3,4}]{Brigitte}{Schmieder}
\author[affil={5,6}]{Reetika}{Joshi}
\author[affil={7}]{Bhuwan}{Joshi}
\author[affil={8}]{Arun Kumar}{Awasthi}
\affiliation[1]{Department of Physics, DSB Campus, Kumaun University, Nainital 263 001, India}
\affiliation[2]{School of Astronomy \& Space Science and Key Laboratory of Modern Astronomy and Astrophysics, Nanjing University, Nanjing 210023, China}
\affiliation[3]{LESIA, Observatoire de Paris, CNRS, 92290 Meudon Principal Cedex, France}
\affiliation[4]{Centre for Mathematical Plasma Astrophysics, Department of Mathematics, KU Leuven, 3001 Leuven, Belgium}
\affiliation[5]{Rosseland Centre for Solar Physics, University of Oslo, N-0315 Oslo, Norway}
\affiliation[6]{Institute of Theoretical Astrophysics, University of Oslo, N-0315 Oslo, Norway}
\affiliation[7]{Udaipur Solar Observatory, Physical Research Laboratory, Udaipur 313004, India}
\affiliation[8]{Space Research Centre, Polish Academy of Sciences, Bartycka 18A, 00-716 Warsaw, Poland}
\correspondance{rchandra.ntl@gmail.com}
\maketitle

%
\vspace{-1cm}
\begin{abstract} 
Extreme-ultraviolet (EUV) waves are one of the large-scale phenomena on the Sun. They are defined as large propagating fronts in the low corona with speeds ranging from a few tens \kms~to a multiple of 1000 \kms. They are often associated with solar filament eruptions, flares, or coronal mass ejections (CMEs). EUV waves show different features, such as, wave and nonwave components, stationary fronts, reflection, refraction, and mode conversion. Apart from these, they can hit the nearby coronal loops and filaments/prominences during their propagation and trigger them to oscillate. These oscillating loops and filaments/prominences enable us to diagnose 
 coronal parameters such as the coronal magnetic field strength. In this article, we present the different observed features of the EUV waves along with existing models.
\end{abstract}


\section{Introduction}
Solar activities can be roughly divided into two categories, namely large and small scales.
Among large-scale phenomena, Moreton waves (in the solar chromosphere) and solar extreme-ultraviolet (EUV) waves are very interesting and important ones. 
Moreton waves were discovered by \citet{Moreton60} in H$\alpha$ centre, blue, and red wings as moving bright and dark fronts, respectively. Their reported speeds are $\sim$ 500 -- 2000 \kms.
EUV waves are defined as large propagating bright fronts clearly visible in the low corona, almost in all directions. They were discovered by the EUV imaging telescope \cite[EIT;][]{Delaboudiniere95} onboard Solar and Heliospheric Observatory \citep[SOHO;][]{Domingo95} and named as the EIT waves. The first reported case study of EIT waves is the 1997 May 12 event which was investigated by \citet{Thompson98}. 
With the better spatio-temporal resolution observations by Solar Dynamics Observatory \citep[SDO;][]{Pesnell12}, there are more observations of EUV waves
\citep[for example,][]{Zhukov04, Chen2011, Chandra2021, Chandra22, Devi2022}. 
Multi-viewpoint observations of Solar TErrestrial RElations Observatory \citep[STEREO;][]{Kaiser08} twin satellite provide an opportunity to investigate the 3D structure of the phenomenon \citep{Attrill09, Zhukov09, Veronig2010, Warmuth2011, Long2011, Muhr2014, Long17, Tatiana2019}. The detailed description of the EUV waves with multi-wavelength and multi-viewpoint observations are presented in past reviews (for example, \citet{Warmuth15} and \citet{Chen16a}. 
In the past decades, EIT waves were also called EUV waves, coronal waves, solar tsunami, large-scale coronal propagating fronts, etc. According to observations, their reported speeds are $\sim$ 10 to more than 1000 \kms. More discussion on their speed is given in Section \ref{sec3}. For consistency, we call them EUV waves throughout this article. 
In addition to EUV wavelengths, they are also visible in radio wavelengths \citep{Aurass2002, Pick2005, Vrsnak2005, Warmuth15}. For the radio observations, mostly the data of Nobeyama radioheliograph \citep[NoRH;][]{Nakajima1994} and Nan\c{c}ay radioheliograph \citep[NRH;][]{Kerdraon1997} were used. \citet{Vrsnak2005} presented the radio counterparts of the EUV waves using NRH data. They found that the wave fronts are cospatial in EUV, H$\alpha$, and X-rays. The development of the EUV wave observed at different NRH frequencies was also presented by \citet{Pick2005}. The radio signatures of EUV wave were also observed in the microwave with the NoRH dataset at 17 GHz by \citet{Aurass2002} and \citet{Warmuth2004}.

As far as the association of EUV wave with solar flares or coronal mass ejections (CMEs) is concerned, it is believed now that EUV waves are more associated with CMEs. The association between EUV waves and CMEs was initially investigated by \cite{Biesecker02}, and they found a strong correlation between them, while in terms of solar flares this association is weak. Using the data of EIT and Large Angle and Spectroscopic Coronagraph \citep[LASCO;][]{Brueckner95}, \citet{Kay2003} examined 69 ejective and non-ejective flares and found that all EUV wave associated flares are accompanied by CMEs. \citet{Chen06} selected a set of 14 non-CME associated flares. They selected energetic flares as they are excepted to generate stronger pressure pulses. It was found that none of the selected flares are associated with EUV waves. \citet{Chen09} examined an EUV wave and its association with CME using the data of EIT and the high-cadence Mark-III K-Coronagraph (MK3) at Mauna Loa Solar Observatory \citep[MLSO;][]{Fisher81}. He found that EUV wave fronts and CME leading fronts are well coaligned. With the SDO observational data sets the CME association with EUV waves has been performed by other authors and they found the association rate varies from 65 to 79 $\%$ \citep{Nitta13, Nitta14, Muhr2014}. The minimum 65$\%$ association is from \citet{Nitta13}, who selected only the solar disk EUV waves.

In the following sections, we present an overview of the existing EUV wave models and their different observational evidence. The paper is organized as follows: Section \ref{sec2} presents a brief summary of existing models. Different observational features are described in Section \ref{sec3}. The use of EUV waves for coronal seismology is given in Section \ref{sec4}. Finally, a short summary is presented in Section \ref{sec6}.

\vspace{-0.5cm}
\section{Existing Models}
\label{sec2}
Since the discovery of EUV waves, several models have been proposed by different investigators. The main models include wave, nonwave, and hybrid models, which are described in brief as follows:

\noindent \textbf{Wave Model:}
    Initially, EUV waves (upon the discovery, they were known as EIT waves) were assumed to be fast--mode MHD waves or shock waves \citep{Thompson1999,Wang2000,Wu2001,Warmuth2001,Ofman2002}. It was believed that they are coronal counterparts of Moreton waves \citep[for example, ][] {Asai12}. \citet{Uchida68} developed  a numerical MHD model to explain Moreton waves. According to this model, 
    the shock wave is generated by the high pressure pulse in the flaring loops. It was later pointed out that the shock wave may not be due to the pressure pulse, and should be piston-driven by an erupting filament or CME \citep{Chen02}.
    Apart from the fast--mode wave model, the slow-mode soliton model \citep{ WillsDavey2007} and Magneto-acoustic surface gravity waves \citep{ Ballai2011} were also proposed. The observational features, such as reflection, transmission, refraction, and mode conversion, tend to support that there is a wave component in EUV waves as presented in Section 3 of this article.

\noindent \textbf{Non-Wave Model:}
After the discovery of EUV waves, a lot of studies have been performed using various instruments all over the globe. In particular, the work on EUV waves became more elaborated after the launch of Atmospheric Imaging Assembly \citep[AIA;][]{Lemen12} onboard SDO. People reported many peculiar features of this interesting phenomenon. Examining the temporal evolution of EUV waves, it is found that the estimated values of speed with the manual tracking as well as time-distance techniques vary from tens to more than 1000 \kms~\citep[][and references therein]{Thompson00, Zhukov04, Chen09, Nitta13, Chandra2018, Chandra2021}. For the first time \citet{Delannee99} reported stationary fronts associated with the EUV waves using the EIT instrument data. They also noticed that this stationary brightening is co-spatial with a magnetic quasi-separatrix layer (QSL). The very low speed of EUV waves together with the reported stationary fronts is the main reason to doubt the wave nature of EUV waves. 
To explain the stationary fronts of EUV waves, \citet{Delannee99} put forward the idea of nonwave model (also known as the magnetic reconfiguration model). According to it the consequence of the reconfiguration of the magnetic field is due to the eruption of CMEs. They conjectured that an EUV wave is the disk projection of the expanding CME. Based on further 3D MHD simulations, \citet{Delannee2000} proposed a current shell model to explain the EUV waves. In their simulations, they found that due to an erupting flux rope, a current shell is formed around it and because of the Joule heating of the current shell, the EUV wave is observed.
A successive reconnection model was also proposed to explain EUV waves \citep{Attrill07, vanDrielGesztelyi2008, Cohen2009, Cohen2010}. According to this model, the EUV wave is a result of reconnection between the expanding CME and quiet magnetic loops \citep[see Figure 4 of][]{Attrill07}. 

\noindent \textbf{Hybrid Model:}
On the one hand, solar flares/CMEs can definitely drive fast-mode waves; on the other hand, multi-wavelength and multi-vantage observations revealed many characteristics in EUV waves that cannot be accounted for by any wave model \citep{Delannee99, Warmuth2004, Balasubramaniam2005, Attrill07, Del2007}.
Keeping this in mind, \citet{Chen02,Chen05} performed MHD numerical simulations of flux rope eruptions (see their Figures 7 $\&$ 8). They found that after a flux rope erupts, two wavelike phenomena with different speeds are observed in the solar corona. The faster wave is a piston-driven shock wave propagating ahead of the erupting flux rope. The leg of this wave travels outward in the horizontal direction and was explained as the coronal Moreton wave, i.e., the wave component of the EUV waves. The slower wavelike features also propagate outward but behind the faster component of the EUV waves. They claimed that the slower component corresponds to the EIT wave observed first time by the EIT onboard the SOHO satellite. 
To explain the formation of the slower component of the EUV waves, they proposed a hybrid model, i.e., an erupting flux rope would generate two types of EUV waves, or there are two components of EUV waves. The faster one is a fast-mode MHD wave or shock wave and the slower component, i.e., the nonwave component, is generated due to the successive stretching of magnetic field lines straddling over the erupting flux rope.
Since this model explains both the wave and the nonwave components of the EUV waves, it is known as a hybrid model.
Very recently, \citet{Guo2023} performed a 3D data-driven radiation MHD simulation of the 2021 October 28 EUV wave event, where they confirmed the coexistence of two components of EUV waves predicted by the magnetic stretching model. They also verified the cospatiality between the CME piston-driven shock and the fast EUV wave component together with the cospatiality between the CME leading front and the nonwave component of the EUV waves.

\vspace{-0.5cm}
\section{Observational Features}
\label{sec3}
The observational features of EUV waves are explained as follows:

\noindent\textbf{Two Components:}
EUV waves were discovered by the EIT instrument onboard SOHO satellite and people calculated their speeds. The first reported EUV wave event on 1997 May 12 was studied by \citet{Thompson98} with a lower temporal resolution. By tracking the wave leading edges in different directions, the measured speed of the EUV wave was 245 \kms. Further, the statistical studies on EUV waves using SOHO/EIT and STEREO/EUVI found average speeds of 200 -- 500 \kms~ \citep{Klassen2000,Thompson09, Muhr2014}. On the other hand, some authors found the speeds of some EUV waves to be less than the sound speed in the corona \citep{Tripathi2007, Thompson09} and in some cases it is only $\sim$ 10 \kms~\citep{Zhukov2009}. These observations actually imply the existence of two types of EUV waves \citep{Chen16a}. For the first time, the observations of two components of EUV wave were reported by \citet{Harra03} using the better time resolution (1 to 2 min) data of the Transition Region and Coronal Explorer \citep[TRACE;][]{Handy1999} satellite. They reported that the faster and slower component front speeds are $\sim$ 500 \kms~and $\sim$ 200 \kms, respectively. They also reported that the faster front is fainter than the slower front.

\begin{figure} [t]
\centering
\includegraphics[width=0.6\textwidth]{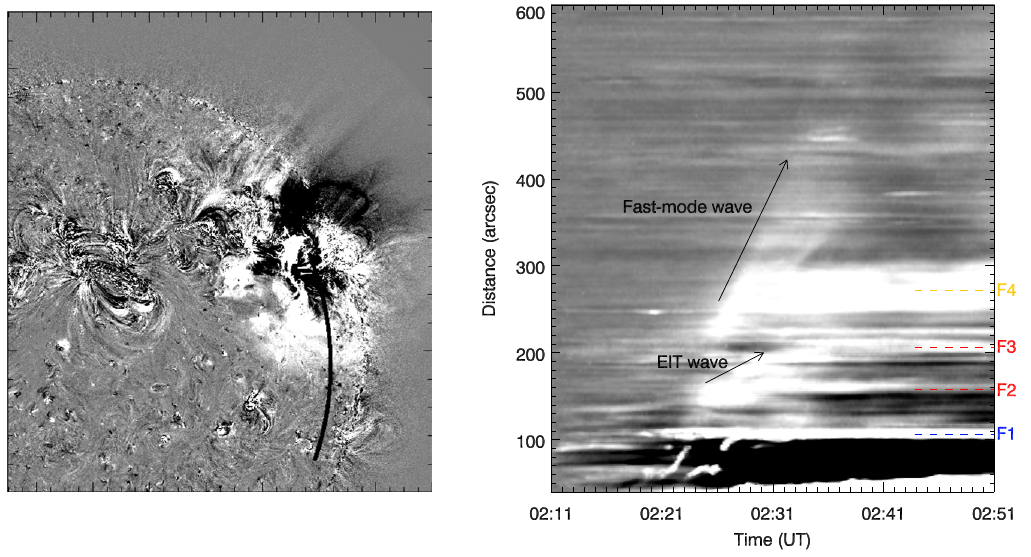}
\bigskip
\begin{minipage}{12cm}
\vspace{-0.5cm}
\caption{Left: AIA 193 \AA\ running difference image on 2011 May 11 at 02:11 UT. 
Right: The time-distance plot along the curved  slice and the location of fast, non-wave component along with stationary fronts: F1 -- F4.
\citep[adapted from][]{Chandra2016}.}
\label{fig1}
\end{minipage}
\end{figure}

Due to the low temporal resolutions of the earlier observations, the two components of EUV waves, which were predicted by the hybrid model, can not be distinguished clearly. It is possible that due to the high speed of the fast mode wave component, it has already travelled out of the field of view (FOV) of observing instruments. Before being observed with the high spatio-temporal resolution SDO data,
several events were analysed and the speeds of the EUV waves were calculated with the time-distance technique. Many of the studies evidenced the two components of EUV waves \citep{Chen11,Asai12,White13, Guo15}.
However, some EUV wave events do not show both components together \citep{Nitta13, Hou2022, Wang2022, Zheng2022}.
An example of the existence of two components of EUV waves is displayed in Figure \ref{fig1}. The faster component of the EUV waves is a real MHD wave while the slower component is the nonwave component (or previously reported EIT wave). According to the hybrid model, the faster front is interpreted as a fast-mode MHD wave or shock wave and the inner slower component corresponds to plasma compression due to successive stretching of magnetic field lines which are pushed by an erupting flux rope. Using the two view-point observations of AIA and STEREO--B instruments, \citet{Chandra2021} confirmed the existence of the fast-mode and nonwave components of EUV waves. They found that the location of nonwave component spatially coincides with the nonwave component observed by STEREO--B.
Regarding the speeds of wave and nonwave components of the EUV waves, \citet{Chen16a} presented excellent discussions. According to him, the wave whose speed is greater than 500 \kms~is a fast-mode wave and that less than 300 \kms~is nonwave in nature. If the speed is between these two limits, i.e., 300 to 500 \kms, it is difficult to determine the nature of the wave. In this case, the nature of wave depends upon other kinematics properties such as whether it stops near the QSLs, and its refraction/reflection when encountering magnetic features.

\begin{figure}[t]
\centering
\includegraphics[width=0.5\textwidth]{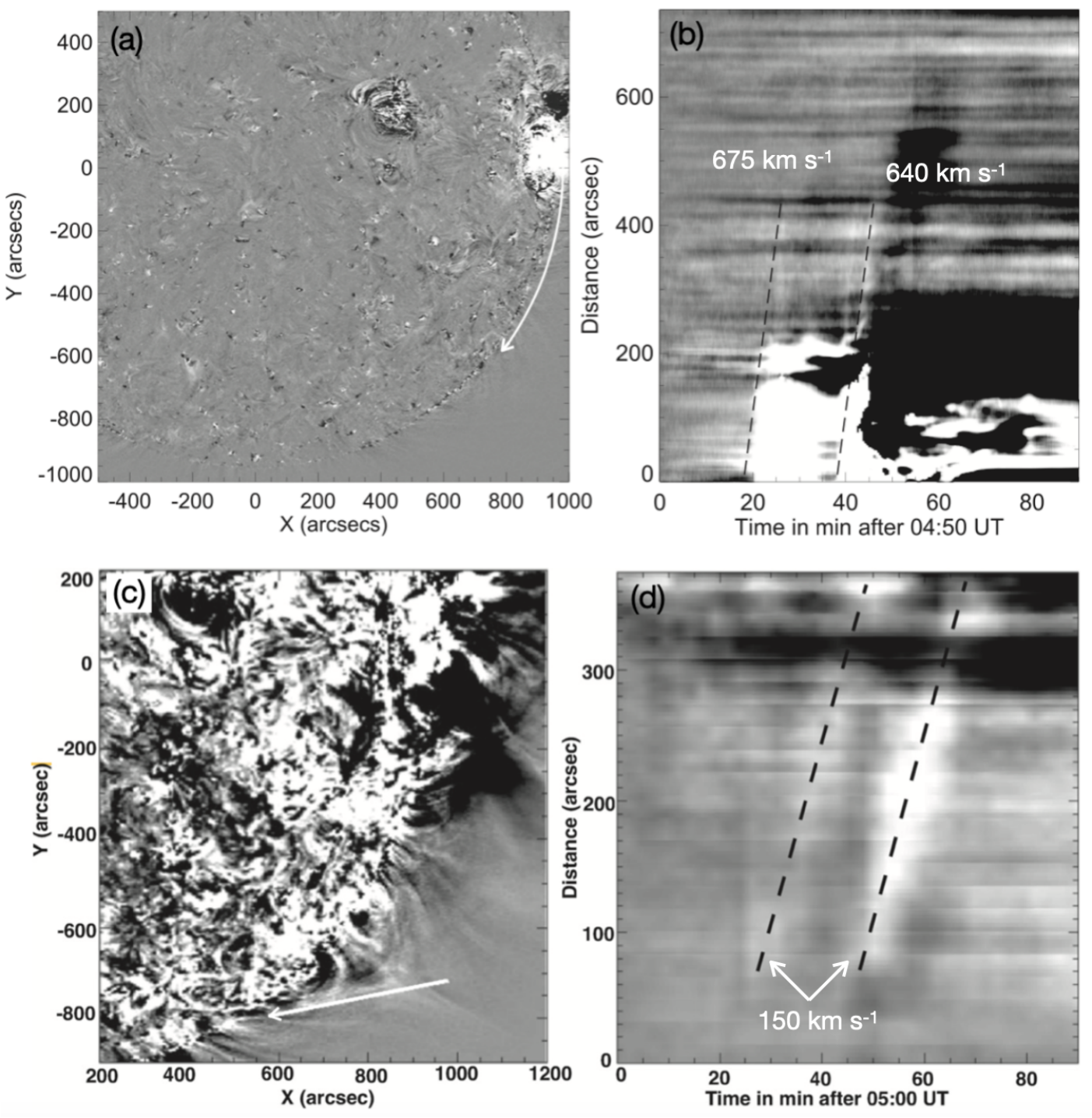}
\bigskip
\begin{minipage}{10cm}
\vspace{-0.3cm}
\caption{EUV wave mode conversion through helmet streamer on 2016 July 23. 
The locations of slits and corresponding time-distance plots  are shown in the figure. The images shown in (a) and (c) are AIA 193 \AA\ at 05:12 and 05:59 UT.
\citep[adapted from][]{Chandra2018}.}
\label{fig2}
\end{minipage}
\end{figure}

\noindent\textbf{Stationary Fronts:}
 \citet{Delannee99} for the first time reported the existence of brightening for several hours in the same location. This brightening is now well known as stationary brightening. Further, \citet{Delannee2000} extended their study and reported more cases of stationary brightening. Using the high temporal and spatial resolution data of AIA onboard SDO,  \cite{Chen11} analysed the EUV wave event of 2010 July 27 and presented the temporal evolution of wave with time-distance diagram along the selected artificial slices. They also observed the stationary front associated with the nonwave component in the time-distance diagram located 250$''$ from the flare site. They investigated the magnetic topology of the stationary front and identified a magnetic separatrix at that location. \citet{Del08} explained the stationary fronts by the current shell model. On the other hand, such stationary brightenings can also be explained by the magnetic field-line stretching model of \citet{Chen02,Chen05}.
 \citet{Chandra2016} analysed the event of 2011 May 11 and reported several stationary fronts. They also compared their locations with the PFSS extrapolated magnetic field and found that their locations are very close to magnetic separatrices, as expected in the magnetic field-line stretching model. Some of the stationary fronts are shown in Figure \ref{fig2}.

\noindent\textbf{Mode Conversion:}
As mentioned in the above subsection, the stationary fronts were initially observed as the final position of the nonwave component of the EUV waves. Moreover, for the first time
 \citet{Chandra2016} analyzed an EUV wave event and reported that together with the nonwave component, a fast-mode component of the EUV waves also produced a stationary front close to a QSL. Based on their observations, they tentatively proposed a wave-trapping model. According to their interpretation, as a fast-mode wave propagates across a magnetic QSL, part of the fast-mode wave is trapped inside the cavity and part of it moves ahead. 
 It is well known that when a fast-mode wave penetrates into the site of weak magnetic field (where the Alfv\'en speed is comparable to the
 sound speed), a part of the fast-mode wave converts into a slow-mode wave \citep{Cally05}. Such a mode conversion can also happen in solar coronal conditions.
 Keeping this fact in mind and motivated by the observational features reported by \citet{Chandra2016}, \citet{Chen16b} did numerical simulations of the interaction between a fast-mode wave and a magnetic QSL. 
 In their simulations, it is revealed that when the fast-mode shock wave enters into a region with weak magnetic field around a QSL, partially it is converted to a slow-mode wave and afterward the slow-mode wave travels along the magnetic field lines with the local sound speed. Finally, the slow-mode wave stops at the location in front of the magnetic separatrix.
Later, the observation of stationary fronts at magnetic QSLs were found by \citet{Fulara19}. They found that the fast-mode component of the EUV waves encounters two QSLs and at both QSLs locations stationary fronts are observed (see their Figure 11). 

Afterwards, more and more examples of mode conversion were reported in EUV wave events \citep{Zong17, Chandra2018, Zheng18}. In \citet{Zong17}, the fast-mode wave interacts with the coronal cavity and after the interaction, it converts into a slow-mode wave. According to \citet{Chandra2018}, two fast-component EUV waves originated from two filament eruptions and both were converted into slow-mode waves. 
 \citet{Zheng18} also found the mode conversion when fast--mode EUV waves interact with coronal streamers. They named it as `secondary wave'.
 Here, we would like to mention that the tips of the helmet streamers and coronal cavity map to magnetic QSLs, and their low Alfv\'en speeds favor the mode conversion. One example of EUV wave mode conversions at the helmet streamers is presented in Figure \ref{fig2}. 

\noindent\textbf{Reflection and Refraction:}
Reflection and refraction are strong evidence for the fast-mode wave component in EUV waves. Reflection happens around coronal holes (CHs), active regions (ARs), and helmet streamers
\citep{Long08,Veronig08,Gopalswamy09}. This phenomenon was discovered using the STEREO data nearly a decade after the discovery of EUV waves. The long delay in reporting the reflection phenomena (which is very common for waves) may be due to the lower cadence of the EIT telescope and was immediately reported after the availability of STEREO having improved  temporal resolution observations. 
This confirms the conjecture that the fast-mode EUV wave component was missed by the low cadence observations as in the case of EIT data sets. \citet{Gopalswamy09} showed the reflection of an EUV wave from the CH using the time-distance diagram. Notably, after the launch of the SDO satellite, a large number of EUV wave reflection cases were reported at various magnetic structures on the solar surface, such as CHs, ARs, and bright points \citep{Li12,Olmedo12, Yang2013, Shen13}. Total reflection was found in the case of 2015 December 22 event by \citet{Zhou22} from the CH boundary. Their observational results showed that the reflection  was a total reflection because the measured incidence and critical angles satisfy the theory of total reflection, i.e., the incident angle is greater than the critical angle. A example of the wave reflection through the CH and AR in a single event of 2011 August 4 was investigated by \citet{Yang2013} and presented in Figure \ref{fig3}. However, it should be noted that not all CHs reflect EUV waves \citep{Chandra22}.
\citet{Thompson00} reported the refraction of the EUV wave from an AR for the first time. After \citet{Thompson00}, the refraction of EUV waves was observed by many other authors \citep{Wills99, Ofman2002, Shen12, Yang2013, Liu2018}.
Using the 2.5D MHD, Piantschitsch and coworkers did numerical simulations of the interactions of a fast--mode MHD wave with CHs and revealed the phenomena of reflection, refraction, and transmission of the wave \citep{Piantschitsch2017, Piantschitsch2018b, Piantschitsch2018}. 
In 3D MHD simulations, \cite{Ofman2002} also reported the reflection, refraction, and dissipation of the wave with small transmission after the interaction of MHD wave with an AR.

\vspace{-0.5cm}
\section{Coronal Seismology}
\label{sec4}

As an EUV wave starts to propagate on/above the solar disk, 
its faster component characterizes the local fast--mode MHD wave speed, hence it can be used to derive the coronal magnetic field. Besides, it can interact with magnetic structures, such as coronal loops and solar filaments, and disturb them. As a result of this, these structures can either oscillate or erupt. If these structures oscillate, they can provide crucial information that can be used to derive the physical parameters of the corona, such as: the magnetic-field strength ($B$), plasma density, transport coefficients, and heating functions, with a technique known as coronal seismology \citep[][]{Uchida1970, Roberts1984, Nakariakov2001, Nakariakov2005} .
\citet{Mann1999} were the first to use this technique for the measurement of $B$. They considered the coronal transient wave to be a fast magnetosonic wave. They derived the Alfv\'en speed ($v_A$) by using the relation 
$V_{wave} = \sqrt{v_A^2+c_s^2}$, where $V_{wave}$ and $c_s$ are the speed of the EUV wave and the coronal sound speed, respectively. $v_A$ is then used to calculate the $B$ by using the formula, 
$B = v_A \sqrt{4\pi \mu m_pN}$ (in the CGS units). Here, $N = \frac{N_e}{0.52}$ \citep{Newk61} denotes the particle number density, $N_e$ is the electron density, $\mu$ = 0.6, and $m_p$ is the mass of proton.
Using the period and length of oscillating loops, the ratio of the densities inside ($n_{in}$) and outside ($n_{ex}$) the loops can be estimated by using the formula given by  
$\frac{n_{in}}{n_{ex}} = \frac{1}{2}\Big(v\frac{P}{L}\Big)-1 $ \citep{Aschwanden2011}, where $v$ is the global fast magneto-acoustic wave speed, and $P$ and $L$ are the period and length of the oscillating loops, respectively.  

\begin{figure}[t]
\centering
\includegraphics[width=0.75\textwidth]{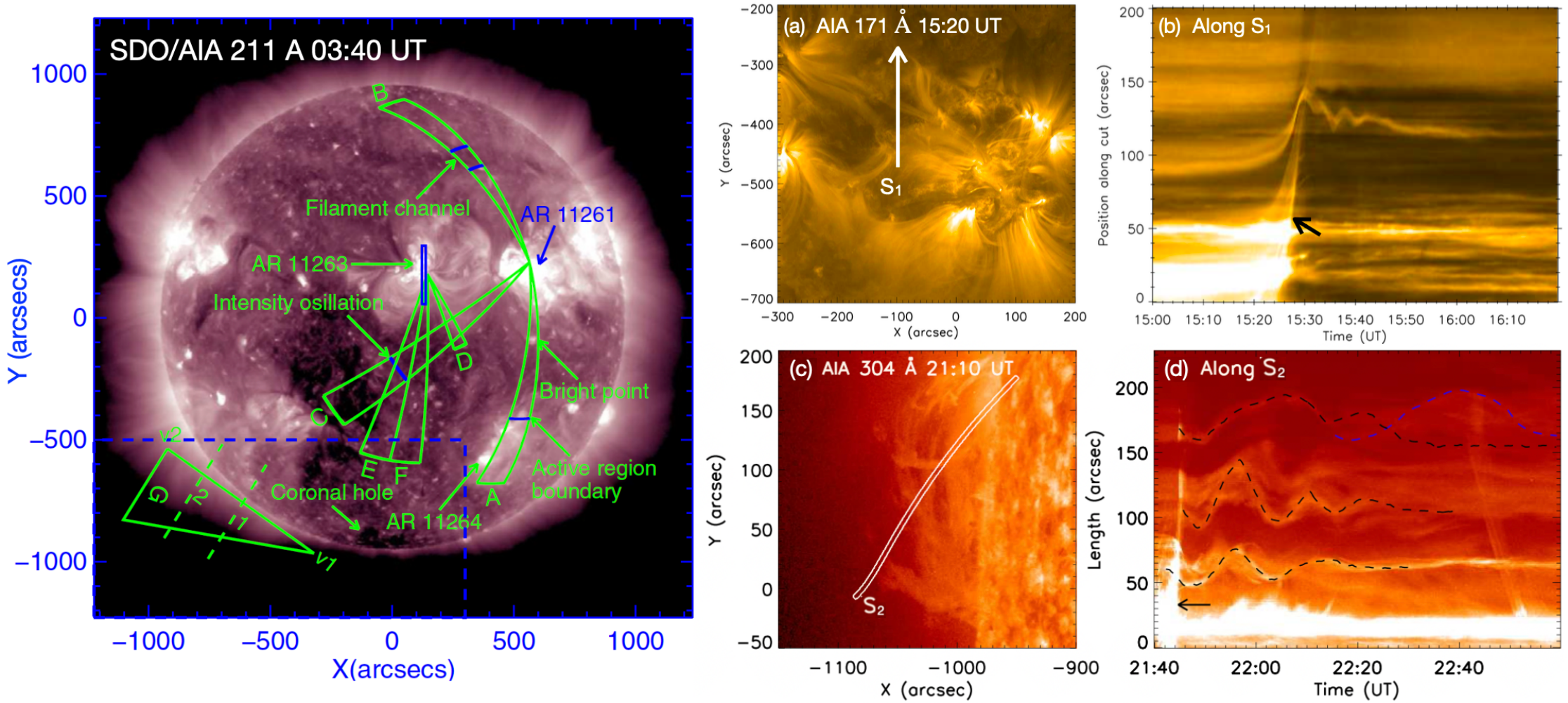}
\bigskip
\begin{minipage}{12cm}
\vspace{-0.3cm}
\caption{
Left: EUV wave reflection though CH and AR  on 2011 August 04 represents by sectors C--F \citep[adapted from][]{Yang2013}.
Right: Loop oscillation on 2021 October 28 with slit S$_1$ and corresponding time-distance plot are shown in (a) and (b). Prominence oscillations created by EUV wave on 2011 February 11 with slit S$_2$ and the time-distance plot are shown in (c) and (d). Black arrows show the EUV waves \citep[adapted from][]{Devi2022sp,Devi2022}. 
}  
\label{fig3}
\end{minipage}
\end{figure}

Using the wave kinematics, several authors derived, the lower coronal magnetic field \citep{Warmuth2005, Ballai07, Devi2022sp}. Mostly their measured values range from 0.5 to 8 G. 
\citet{Devi2022sp} analysed the interaction of EUV wave with neighbouring EUV loops and their oscillations. Their measured coronal magnetic field ranges from 1 to 8 G. In another study, \citet{Devi2022} presented the oscillations of a prominence due to an EUV wave and computed the magnetic field strength in the prominence. The calculated magnetic field value ranges from 14 to 20 G, which is consistent with the previous studies \citep{Mackay2010, Luna2017}. Figure \ref{fig3} displays an example of oscillating EUV loops and filaments along with their time-distance diagrams. It would be interesting to compare the magnetic field strength computed using seismology with radio observational techniques as well as modelling.

\vspace{-0.5cm}
\section{Summary}
\label{sec6}

In this article, we presented a review of the recent observational results of the EUV wave events and modellings. The main points of this review are summarized as follows:

--An EUV wave event is often composed of two components, namely, a fast-mode wave and a nonwave component. Both components can be explained well by the hybrid model.
%

--The fast-mode wave component of the EUV waves was confirmed by its characteristics of reflection, refraction, and mode conversion. The reflection, refraction, and mode conversion were observed at the boundaries of CHs, ARs, and helmet streamers, which often correspond to magnetic QSLs.
%
The observations of stationary brightening, associated with the slower component of EUV waves, evidence the presence of nonwave components in EUV waves.

--Propagating EUV waves may interact with magnetic structures in the solar corona, which may result in oscillations or eruptions of coronal structures. Therefore, it enables tracing the plasma and magnetic field of various coronal structures.

It should be noted that the separation between the wave and nonwave components is very useful and interesting, which can help clarify the association among the EUV waves, type II radio bursts, and solar energetic particle (SEP) events.
Type II radio bursts are created by the shock wave ahead of a CME. Therefore, if EUV waves are the fast-mode shock waves driven by the eruption, EUV waves should be strongly correlated with type II radio bursts. However, many studies found a weak correlation between the speeds of the two phenomena \citep{Klassen2000, Long2017}. The reason for the negative results is that those authors treated the EIT waves, which are the nonwave component of EUV waves, as fast-mode shock waves. Similarly, some authors tried to associate EIT waves with SEPs \citep{Bothmer1997, Torsti1999, Miteva2014}. In our opinion, such kind of studies are meaningful only if we separate the wave and nonwave components of EUV waves. Only the faster component EUV wave can provide correct information associated with type II radio bursts and SEPs.

Here, we would like to mention that more efforts are needed to investigate the relationship between CMEs and EUV waves. For this purpose, 
the CME and EUV observations should have overlapping fields of view as much as possible. 
Ground-based coronagraphs like MLSO MK3, MK4, and currently working MLSO KCor can be very useful.
However, ground-based telescopes suffer from the limited duty cycle in observations. Therefore, we think that the CME observations in the inner corona from space are needed. The recently launched ADITYA-L1 Indian spacecraft with its Visible Line Emission Coronagraph (VELC) instrument can provide important observations to understand these phenomena in more detail.

\begin{acknowledgments}
The authors thank the open data policy of the SDO and STEREO, and SOHO teams. P.F.C. is finally supported by the National Key Research and Development Program of China (2020YFC22- 01200) and NSFC (12127901). P.D. is supported by CSIR, New Delhi. R.J. acknowledge the support by the Research Council of Norway through its centres of Excellence scheme, project number 262622. We acknowledge the BINA conference organized at Nainital, India. We thank the referee for the comments and suggestions.
\end{acknowledgments}

\begin{furtherinformation}

\begin{orcids}
\orcid{0000-0002-3518-5856}{Ramesh}{Chandra}
\orcid{0000-0003-0713-0329}{Pooja}{Devi}
\orcid{0000-0002-7289-642X}{P. F.}{Chen}
\orcid{0000-0003-3364-9183}{Brigitte}{Schmieder}
\orcid{0000-0003-0020-5754}{Reetika}{Joshi}
\orcid{0000-0001-5042-2170}{Bhuwan}{Joshi}
\orcid{0000-0003-1948-1548}{Arun Kumar}{Awasthi}

\end{orcids}

\begin{authorcontributions}
RC, PD, BS, BJ, RJ, and AKA contributed to the data analysis. RC  wrote the main draft of the paper. PFC wrote substantial parts of the manuscript and contributed to the interpretation. All the authors did a careful proofreading of the text and references.
\end{authorcontributions}

\begin{conflictsofinterest}
 The authors declare no conflict of interest.
\end{conflictsofinterest}

\end{furtherinformation}
\bibliographystyle{bullsrsl-en}
\bibliography{references}

\end{document}